\documentclass[10.5pt]{article}

\usepackage{geometry}
\usepackage{authblk}
\usepackage{apacite}
\usepackage{amsfonts}
\usepackage{amsthm}
\usepackage{amsmath}
\usepackage{graphicx}
\usepackage{subfigure}
\usepackage[colorlinks,linkcolor=blue,citecolor=blue]{hyperref}
\usepackage{pifont}
\usepackage{natbib}
\usepackage{booktabs}
\usepackage{threeparttable}
\usepackage{algorithm}
\usepackage{algorithmic}
\usepackage{amssymb}
\usepackage{dsfont}
\usepackage{mathrsfs}
\usepackage{setspace}
\usepackage{amsmath}

\newcommand{\Nor}{\mathcal{N}}
\usepackage{bm}
\newcommand{\bbeta}{\boldsymbol{\beta}}
\newcommand{\bb}{\boldsymbol{b}}
\usepackage{multirow}
\usepackage{longtable}

\graphicspath{ {./Figure/} }

\providecommand{\keywords}[1]{\textbf{\textit{Keywords:}} #1}

\title{\bf{Predication of Inflection Point and Outbreak Size of COVID-19 in New Epicentres}}
\author[a]{\normalsize \bf Qibin Duan}
\author[a]{\normalsize \bf Jinran Wu}
\author[b]{\normalsize \bf Gaojun Wu}
\author[a]{\normalsize \bf You-Gan Wang \thanks{To who all correspondence should be directed, {\it Email: you-gan.wang@qut.edu.au}}}
\affil[a]{\normalsize  \bf  School of Mathematical Sciences, Queensland University of Technology, Brisbane 4001, Australia }
\affil[b]{\normalsize  \bf The First Affiliated Hospital of Wenzhou Medical University, Wenzhou 325000, China }

\begin{document}
\maketitle

\setcounter{page}{1}

\raggedright

\begin{abstract}
\noindent The coronavirus disease 2019 (COVID-19) had caused more that 8 million infections as of middle June 2020. Recently, Brazil has become a new epicentre of COVID-19, while India and African region are potential epicentres. This study aims to predict the inflection point and outbreak size of these new/potential epicentres at the early phase of the epidemics by borrowing information from more “mature” curves from other countries. We modelled the cumulative cases to the well-known sigmoid growth curves to describe the epidemic trends under the mixed-effect models and using the four-parameter logistic model after power transformations. African region is predicted to have the largest total outbreak size of 3.9 million cases (2.2 to 6 million), and the inflection will come around September 13, 2020. Brazil and India are predicted to have a similar final outbreak size of around 2.5 million cases (1.1 to 4.3 million), with the inflection points arriving June 23 and July 26, respectively.  We conclude in Brazil, India, and African the epidemics of COVI19 have not yet passed the inflection points; these regions potentially can take over USA in terms of outbreak size.

\end{abstract}


\section{Introduction}

The severe acute respiratory syndrome coronavirus 2 (SARS-CoV-2), first reported in Wuhan, China at the end of 2019, spread across China and the globe and was declared a pandemic on March 11, 2020. As of middle June 2020, it had caused more than 8 million infections. As reported by \href{https://www.who.int/docs/default-source/coronaviruse/situation-reports/20200708-covid-19-sitrep-170.pdf?sfvrsn=bca86036_2}{WHO (2020)},  in recent months, USA and Europe have been epicentres of COVID-19 and are experiencing the rapid increase of its outbreak size. Although the growth of daily confirmed infections in these regions is slowing down, there is still no sign that the epidemic has gained any control or “flattened”  in these regions; even worse, Brazil, India and African region (all affected countries) have become the new epicentres of COVID-19 with increasing number of confirmed infection every day \citep{lancet2020covid, pearson2020projected}.\\

Mathematical modelling, including statistical modelling, is an important tool to understand and predict the dynamics of new diseases. Since the first identification of COVID-19, different approaches haves been developed to simulate and characterize its dynamics and spread trend \citep{grasselli2020critical, yang2020modified, roosa2020real, petropoulos2020forecasting, hengjian2020nonlinear, perc2020forecasting, sajadi2020temperature,zhang2020predicting}. The classical one is the dynamic infectious disease modeling, with using deterministic ODE models or stochastic individual based models, and such approach allows to incorporate with underlying mechanisms of spread and various risk factors in the simulation of transmission \citep{peng2020epidemic, wynants2020prediction, kucharski2020early,fanelli2020analysis}. This approach is commonly used to identify the crucial transmission parameters and  assess the potential impact of public health interventions. Although prediction of transmission trend can be also achieved, setting up such models needs heavy information of local demography and praxiology that is difficult to obtain accurately. For the purpose of prediction, data-driven (or phenomenological) methods are preferred, i.e., machine learning and statistical modeling \citep{alimadadi2020artificial, ribeiro2020short, benvenuto2020application, ceylan2020estimation, ribeiro2020short,zheng2020predicting,kavadi2020partial}. 

Almost investigated modeling approaches have been developed to characterize the transmission and impact of COVID-19 in the context of a specific country or region, from which this information can be estimated, e.g., forecasting the confirmed cases and deaths in China \citep{gao2020forecasting, al2020optimization}. However, for many regions, such modelling studies are still unavailable now. Knowledge of the inflection (“flattening the curve”) and maximal outbreak size is crucial to reflect the evolving trend of an epidemic; in the case of COVID-19, this information remains unclear but does influence the dates of changes to policy restrictions and the recovery of the global economy. Furthermore, an accurate prediction of such information at the early stage is difficult due to the lack of detailed data on testing availability and reporting/infection processes as well as governmental restrictions. \\

To address the problems, non-linear mixed effect model is used to model the (transformed) daily reported number of cumulative confirmed cases. The data was grouped according to country or region. In the countries at the later stage of the epidemics, the reported number of cumulative confirmed cases all show a sigmoidal shape with respect to time, so we use the four parameters Logistic model (FPLM), a generalization of Logistic growth model, to model the growth patterns. By fitting to such non-linear mixed effect model we can predict the inflection point and final size of outbreak in modeled countries and regions. \\

\section{Data}

The data set was downloaded from European Centre for Disease Prevention and Control, which provides the geographic distribution of COVID-19 cases worldwide, e.g. daily incidence (newly confirmed cases),  cumulative number of confirmed cases, population of each country, etc., from  \href{https://opendata.ecdc.europa.eu/covid19/casedistribution/csv}{European Centre for Disease Prevention and Control}. Note that we select two groups of countries/regions, the first is for the countries in the late stage of outbreak, including Australia, China, France, Italy, Germany, Spain and UK; and countries/regions in the second group are still in the early stage of the outbreak, including USA, Africa (as a whole), Brazil, India and Russia.
The data set is about the officially confirmed and reported cases, which is inevitably inaccurate and under-reported due to the limited coverage of testing, especially in the early period of the outbreak. Also, this study is more interested in the future growth trend and final outbreak size. Hence, early observations were thrown away to enable the model to fit the early period more flexibly.

\section{Growth curves with random coefficients}

Let $y(t)$ be the cumulative numbers at time $t$ and $Y’(t)$ be the derivative function representing the growth rate. If $x_t$ represents the explanatory variables(such as temperature, or behavour changes due to government restrictions) believed to be related to the growth rate, we need to incorporate their effect in the growth model via a link function
\begin{equation}
y’(t) = f(y(t))g(x_ty;\theta) + \epsilon_t,
\end{equation}
where $f(y)$ specifies the growth rate as a function of its current size under the constant environmental condition while
$g$ models how the growth rate might change when the environmental conditions ($x_t$) changes. Here $\sigma(t)$ is the stochastic error of zero mean representing the environmental or measurement perturbation  possibly with heteroscedasticity. More details can be seen in \cite{wang1999estimating}.

The simple linear function of $f$ corresponds to the asymptotic regression model (also known as von Bertalanffy growth curve).
However, we are particularly interested in a sigmoidal curves and the inflection point is of great interest. The well known  curves of this type include logistic (autocatalytic),  Richards and Gompertz \citep{Nonlinear1989}.\\

In this study, we will apply the mixed-effects models assuming each country follows the same curve but different set of parameters. These parameters can be potentially modelled as functions of population size and other attributes. More details can be seen in \cite{pinheiro2006mixed}.
A model for nonlinear mixed-effects can be written as
\begin{equation} \label{eq:nlmem}
 g(y_{ij}) = f(A_i \bbeta + B_i \bb_i, t_{ij})+\epsilon_{ij}, b_i \sim \Nor(0,\Sigma), \epsilon_{ij} \sim \Nor(0,\sigma^2)
\end{equation}
for observation $j, j=1,\ldots,n_i$ in group $i,i=1,\ldots,M$. In model~\eqref{eq:nlmem}, $\bbeta_i$ includes both fixed effect $\bbeta$ and random effects $\bb_i$. Specifically in our case, $y$ is the cumulative number of confirmed cases, $g$ is the transformation used, $i$ is the index of country/region, and $t$ is time index of the observation (day). Here $A_i$ and $B_i$ are design matrices for $i$th group to determine the fixed and random effects. The advantage of mixed-effect model is to produce more precise estimate of  $\bb_i$ and $\bbeta_i$ by borrowing strength/information from the rest of the sample from the population. See \cite{pinheiro2006mixed} for more detains about non-linear mixed effect model.

To model the confirmed cases over time we use a four Four Parameter Logistic Model (FPLM)
\begin{equation}\label{eq:fpl}
f(\beta_i,t_{ij}) = \phi_{i1} + \frac{\phi_{i2} - \phi_{i1}}{1+\exp\left(\frac{\phi_{i3} - t_{ij}}{\phi_{i4}} \right)},
\end{equation}
where $\beta_i=(\phi_{i1},\phi_{i2},\phi_{i3},\phi_{i4}) $, and the parameters are:
\begin{itemize}
\item $\phi_{i1}$, the minimum theoretical value of $y_{ij}$ as time $t_{ij}\rightarrow -\infty$;
\item $\phi_{i2}$, the maximum value  as $t_{ij} \rightarrow \infty$;
\item $\phi_{i3}$, the inflection point, and  $t_{ij}=\phi_{i3}$ the response $y_{ij}$ is midway between the $\phi_{i1}$ and  $\phi_{i2}$;
\item $\phi_{i4}$, is a scale parameter for time.
\end{itemize}

We are particularly interested in the following two parameters. $n_{max}$: the maximum number (asymptote), and $n_{Infl}$: the number of cases at the inflection point ($\phi_1+\phi_2)/2$.

We first tried this on the raw data (Model 1). We then used power and logarithmic transformation of cumulative number of confirmed cases as the fitting response as follows.

\begin{enumerate}
\item[a.] Power (square root) transformation (Model 2)
\begin{equation}\label{eq.fplpw}
[y_{ij}]^{1/2}  = f(\beta_i,t_{ij}) = \phi_{i1} + \frac{\phi_{i2} - \phi_{i1}}{1+\exp\left(\frac{\phi_{i3} - t_{ij}}{\phi_{i4}} \right)}
\end{equation}
with $n_{max} =\phi_2^2$. $n_{Infl} =\sqrt{n_0 n_{\max}}$ where $n_0=\phi_1^2$;
\item[b.] Logarithmic Transformation (Model 3)
\begin{equation}\label{eq.fpllog}
\log_{10}[y_{ij}] = f(\beta_i,t_{ij}) = \phi_{i1} + \frac{\phi_{i2} - \phi_{i1}}{1+\exp\left(\frac{\phi_{i3} - t_{ij}}{\phi_{i4}} \right)}.
\end{equation}
\end{enumerate}
Note that the parameters of FPLM model have different interpretations under transformations. The models are validated by comparing the first-order difference of modeled $y_{ij}$ to the corresponding reported daily new confirmed cases.

\section{Results}

Here we fit the cumulative cases after power transformation to the well-known growth curves to describe the epidemic trends in the countries and regions of interest under the mixed-effect model and using the four-parameter logistic model. The advantage of the mixed-effect model is to “borrow information” from the members with rich information \citep{pinheiro2006mixed}. The four-parameter logistic model has been proved to perform well in describing epidemic growths \citep{wu2020generalized, chen2020reconstructing}.\\

We included 12 countries and regions in our study. Nine countries (Australia, China, France, Germany, Italy, Russia, Spain, UK and USA) have almost experienced a full growth curve, and other regions (Brazil, India and the African region) at the early phase of the epidemic will have certain similarity with some of these nine countries in transmission and response strategies. This may result in similarity in the growth of confirmed cases. \\

\begin{figure}[H]
	\makebox[\textwidth][c]{\includegraphics[width=1.1\textwidth]{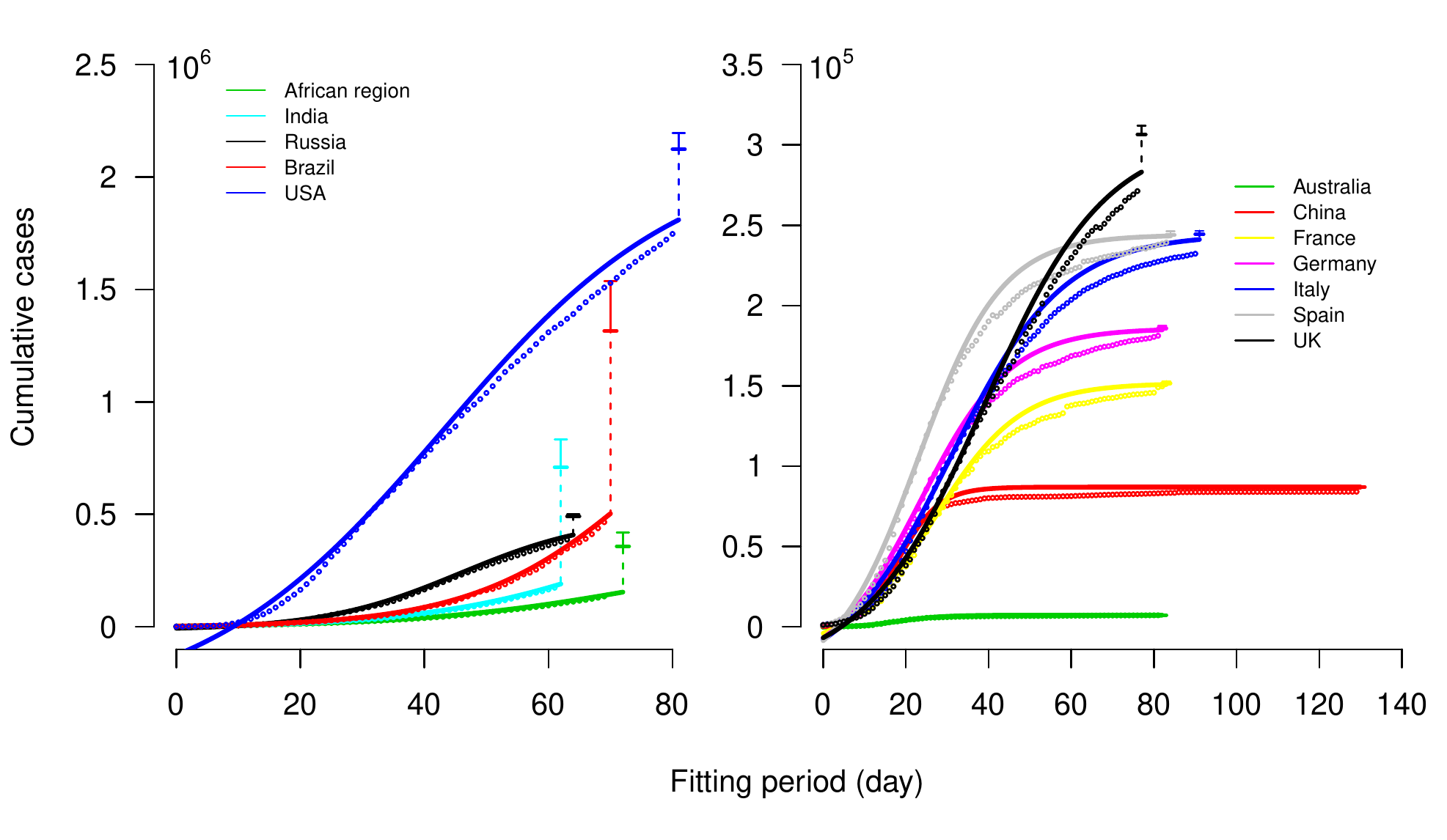}}%
	\caption{Fitting results of model 1 with no transform on cumulative cases.}
	\label{fig:fitm1}
\end{figure}

\begin{figure}[H]	
	\makebox[\textwidth][c]{\includegraphics[width=1.0\textwidth]{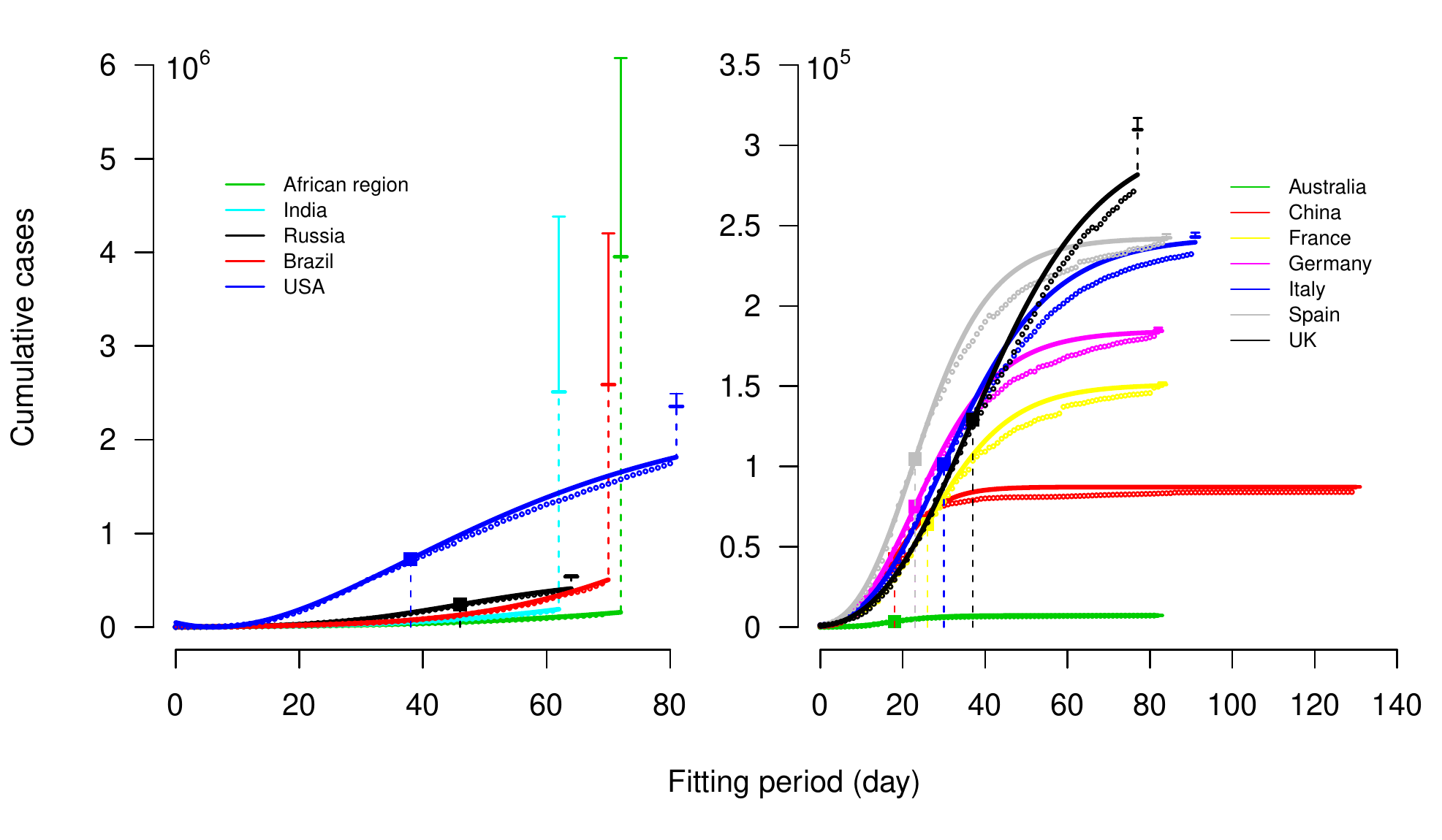}}%
   \caption{Fitting results of model 2 with squared root transform on cumulative cases}
   \label{fig:fitm2}
\end{figure}

\begin{figure}[H]		
	\makebox[\textwidth][c]{\includegraphics[width=1.0\textwidth]{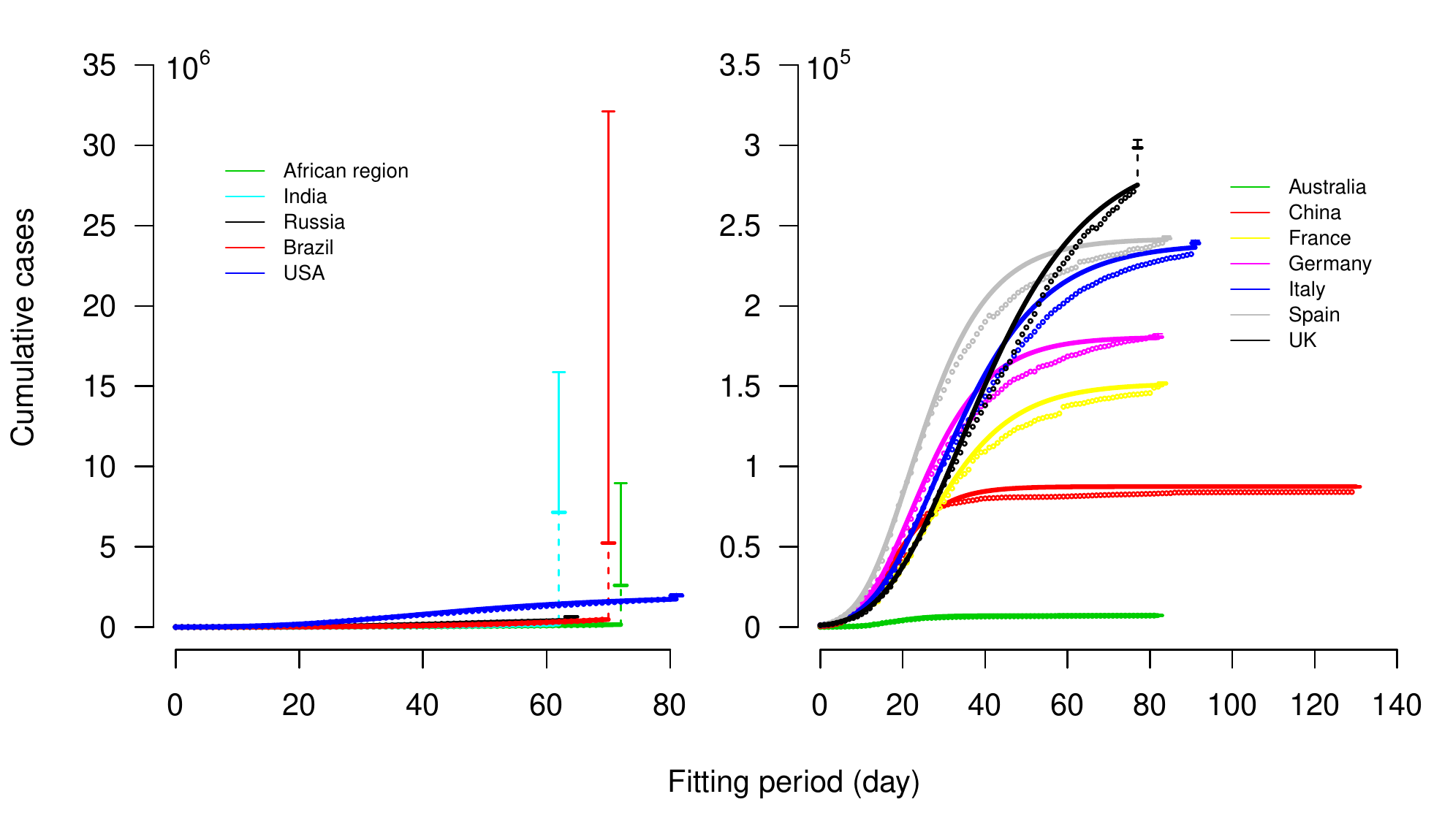}}%
	\caption{Fitting results of model 3 with $\log_{10}$ transform on cumulative cases}
	\label{fig:fitm3}
\end{figure}

\begin{figure}[H]
	\makebox[\textwidth][c]{\includegraphics[width=1.0\textwidth]{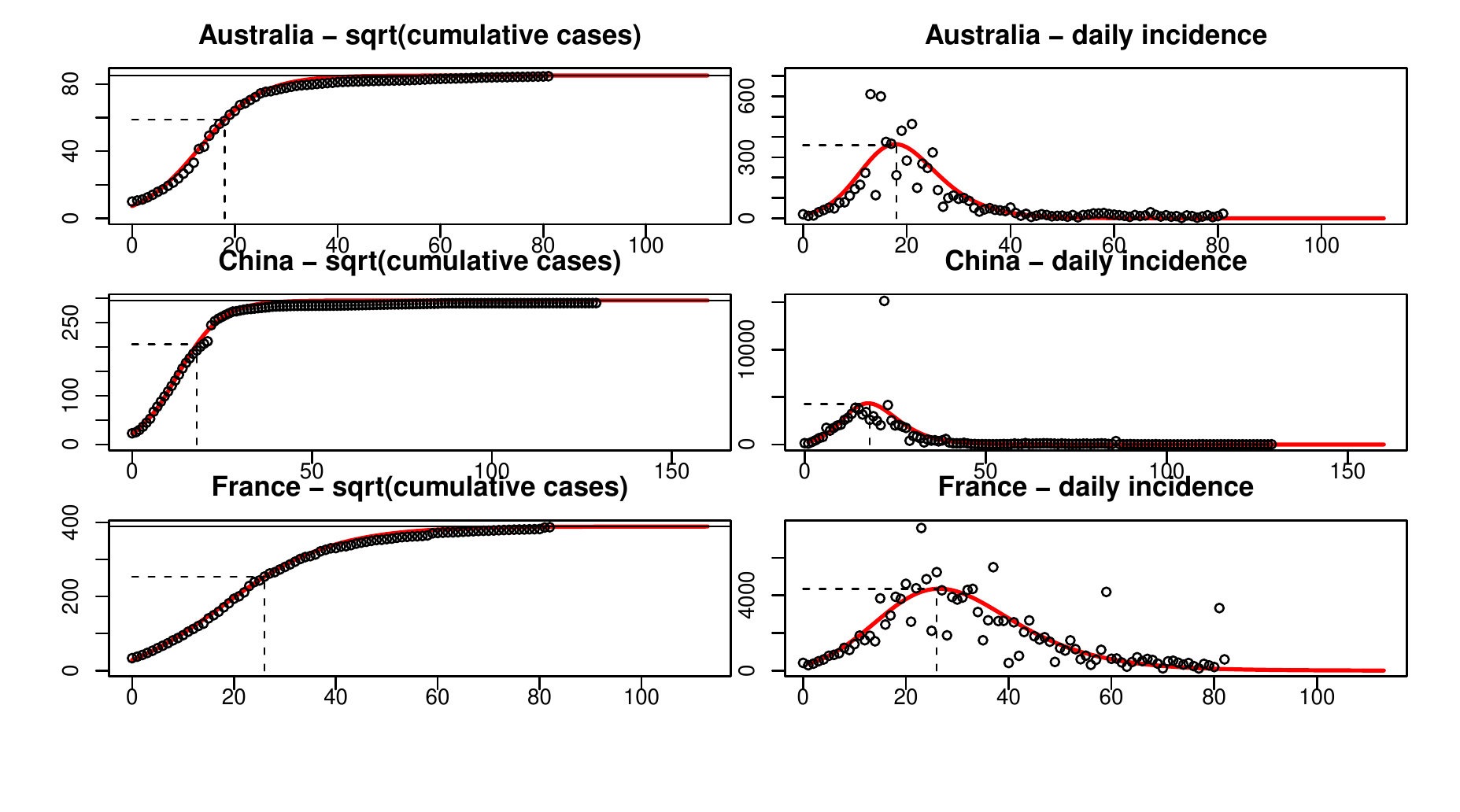}}\\
		\makebox[\textwidth][c]{\includegraphics[width=1.0\textwidth]{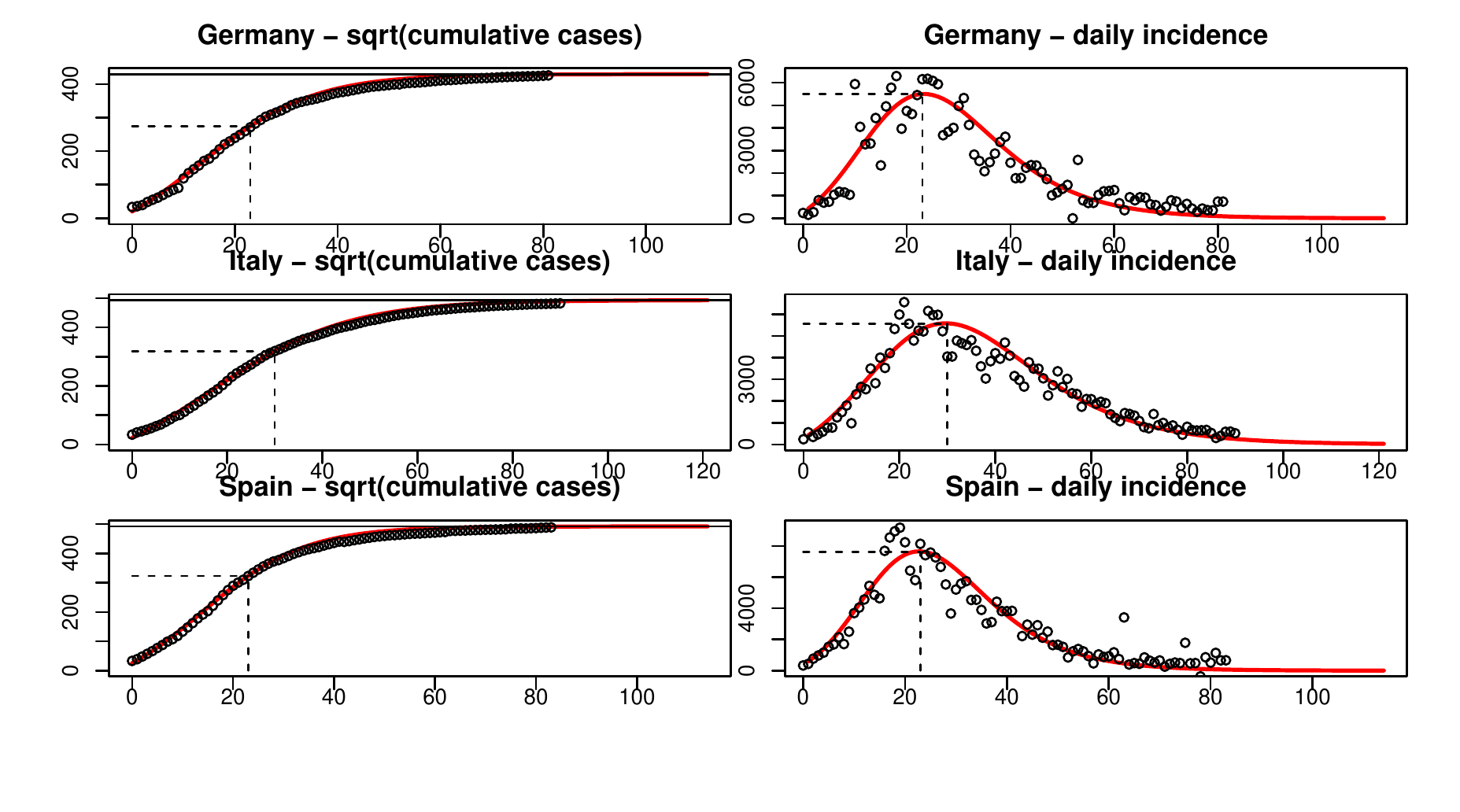}}%
	\caption{Model validation. The left panel of each country shows the squared root of cumulative cases against time, where the black dots are the observed data and red curve is fitted model. The right panel show the daily incidence of each country.  }
	\label{fig:vld1}
\end{figure}

\begin{figure}[H]
	\makebox[\textwidth][c]{\includegraphics[width=1.0\textwidth]{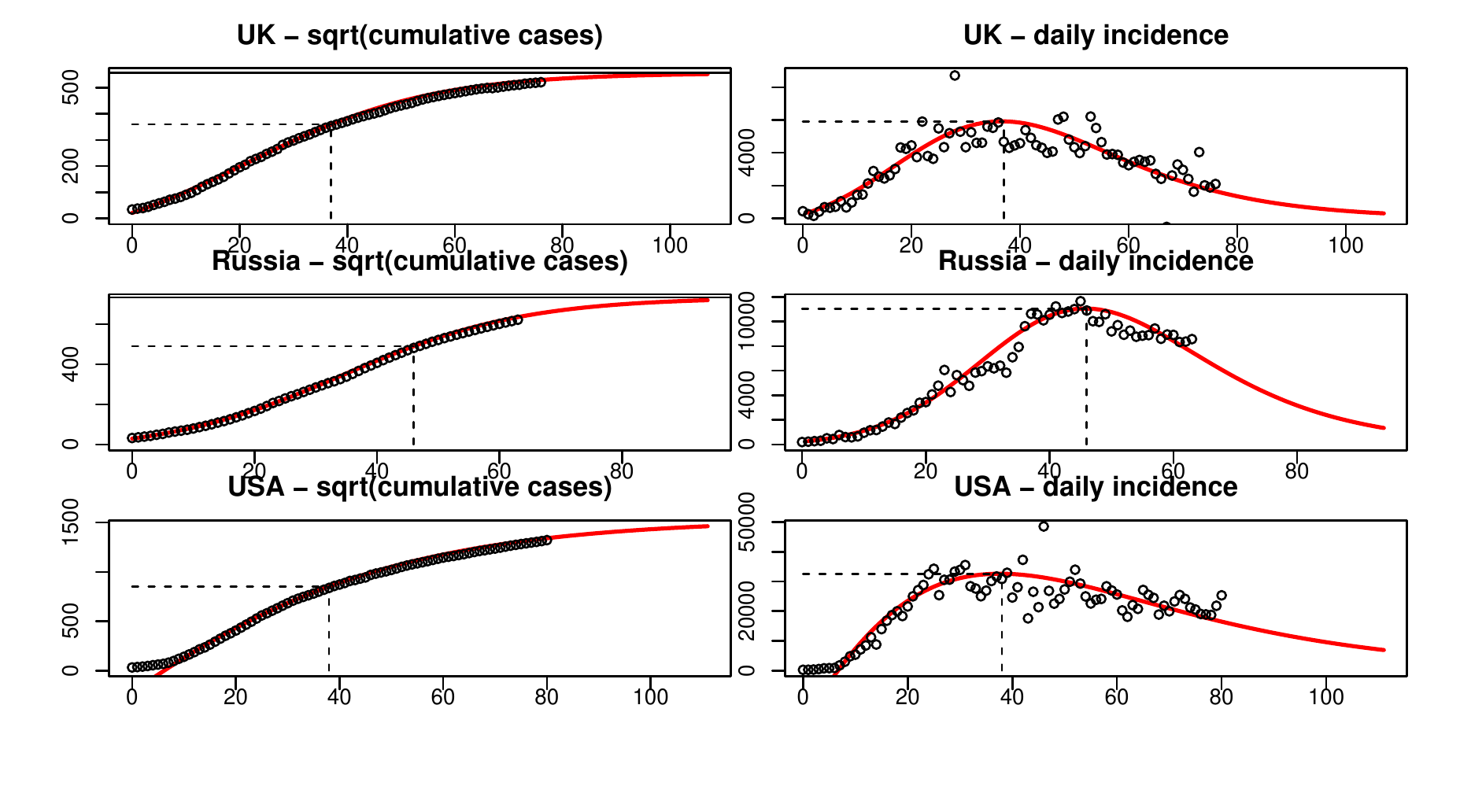}}\\
		\makebox[\textwidth][c]{\includegraphics[width=1.0\textwidth]{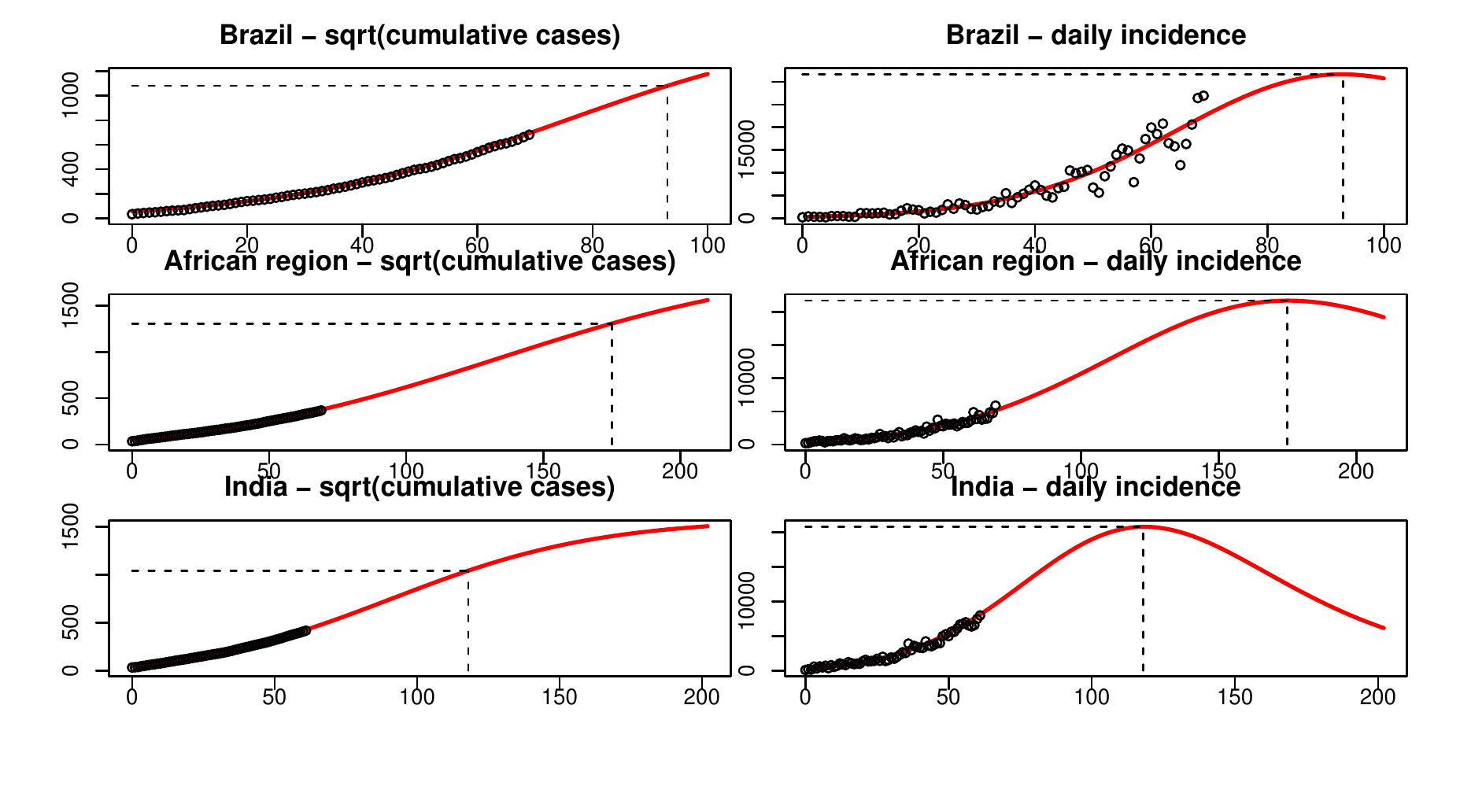}}%
	\caption{Model validation (continued). The left panel of each country shows the squared root of cumulative cases against time, where the black dots are the observed data and red curve is fitted model. The right panel show the daily incidence of each country.}
	\label{fig:vld2}
\end{figure}

The fitted results for all selected countries/regions under three different models are shown in Table~\ref{tab:results}. Note that  $n_0= 10^{\phi_1}$, $n_{max} =10^{\phi_2}$, and $n_{Infl}$ is the number of cases at the inflection point, which is $\sqrt{10^{\phi_1+\phi_2}}=\sqrt{n_0 n_{max}}$. Model 1, 2 and 3 correspond to no transformation, power transformation and log transformation of the cumulative number of confirmed cases in ~\eqref{eq:fpl},~\eqref{eq.fplpw} and~\eqref{eq.fpllog}, respectively. Additionally, Figure~\ref{fig:fitm1}, ~\ref{fig:fitm2} and ~\ref{fig:fitm3} show the fitted curves and data (circles  in the figures) used for fitting of each country or region. (Each color represents the growth curve of one country or region. The dots in the figure are the reported number of cumulative confirmed cases and the curves are the estimated growth curves. The bottom horizontal line segments are the estimate of maximal outbreak size, while top horizontal line segments are the upper boundary of 95\% confidence interval.) Moreover, The reported number of daily confirmed cases (daily incidence in Figure~\ref{fig:vld1} and \ref{fig:vld2}) is used to validate the fitting model.\\

Additionally, here it is statistically sensible to consider a power transformation, $y^\theta$. The well known Box-Cox transformation, $(y^\theta-1)/\theta$, would be a sensible choice (the limiting case of $\theta \rightarrow 0$ corresponds to the log-transformation. An 'optimal' or estimate of $\theta$ can also be obtained via likelihood or other robust statistical approaches. We have chosen three different $\theta$ values (1, 0.5 and 0) in this analysis. It is interesting to see how robust the results are in terms of ballpark estimates of the inflection point and maximum number of cases.\\

Our models have confirmed that the nine countries aforementioned had passed the inflection point (the squared marks in Figure ~\ref{fig:fitm1}). Specifically, China had the inflection point on February 9, 2020, with a final outbreak size of 87k \citep{gu2020inflection}. Australia, Italy, Spain, Germany and France have passed their inflection point in later March and early April; in these countries the current outbreak size almost reaches the estimates of maximal level (with invisible upper confidence boundaries in Figure ~\ref{fig:fitm1}).  Unfortunately, cases in other countries (i.e., USA, Russia, and UK) will continue to increase, possibly to 2,357k (up to 2,425k); 538k (up to 546k); and 309k (up to 313k), respectively. \\

In terms of other three regions (Brazil, India, and African region), they are still in the early outbreak phase (before inflection). Prediction of inflection point with current data for these regions is very difficult, based on the “shape” from other countries, the non-linear mixed effect model provides sensible predictions albeit large error intervals (in the left panel of Figure ~\ref{fig:fitm1}). African region is predicted to have the largest total outbreak size of 3.9 million cases (2.2 to 6 million), and the inflection will come around September 13, 2020. Note that African region here includes all the African countries affected by COVID-19. Brazil and India are predicted to have a similar final outbreak size of around 2.5 million cases (1.1 to 4.3 million), with the inflection points arriving June 23 and July 26, respectively.  The epidemic in Brazil has entered the rapid growth stage and is increasing quickly in the number of the cumulative confirmed cases. The epidemic in India has the similar situation with Brazil, but the growth is predicted to have one-month delay. 
  
\section{Conclusions}

In these developing areas, community transmission and spread of COVID-19 has been ongoing for a while, but large-scale testing is only available until recent weeks. The rapid increase in confirmed cases can be attributed to the increasing level of testing coverage. Currently African region has a relatively small number of confirmed infection ($<$150k), but it might continue to increase to the level estimated in this study if not controlled effectively \citep{pearson2020projected}.\\

Although USA passed the inflection point around April 18,2020, its growth rate (or the number of daily new confirmed infections) failed to achieve rapid decrease. Russia has a similar epidemic curve to USA. This indicates that there are no effective intervention strategies in these countries to further curb the ongoing transmission of COVID-19. Form the validation figures in supplementary material, our model tends to underpredict the daily new confirmed infections in USA and Russia during the later phase of outbreak after inflection points; thus, USA and Russia may have a larger outbreak size than our estimates. Likewise, if these new epicentres (i.e., Brazil, India and African region) could not conduct effective measures to mitigate the spread and transmission, there will be a large number of confirmed infections even after the inflection points. In summary, our model predicted the inflection point and maximal outbreak size in Brazil, India, and African region; these regions might take over USA in terms of outbreak size at the end. \\

This work only fits the growth curves to reported number of confirmed infections, and incorporating localized intervention policies and behavior parameters will improve the performance of fitting and prediction. Furthermore, in this work we test three different power transformation of cumulative number of cases, however, they are likely not the optimal choice for all countries and regions. Other power transformations are worthwhile to test and possibility of using different transformation to different countries/regions should be explored under the same framework of mixed-effect model.   
\\

\section*{CRediT authorship contribution statement}

{\bf Qibin Duan:} Conceptualization, Methodology, Formal analysis, Validation, Writing - original draft, Writing - review \& editing. {\bf Wu:} Methodology, Writing - review \& editing. {\bf Jinran Wu:} Writing - review \& editing. {\bf You-Gan Wang:} Conceptualization, Methodology, Writing - original draft, Writing - review \& editing.

\section*{Declaration of Competing Interest}

The authors declare that they have no known competing financial interests or personal relationships that could have appeared to influence the work reported in this paper.

\section*{Acknowledgement}

This work was supported by supported by the Australian Research Council Centre of Excellence for Mathematical and Statistical Frontiers (ACEMS), under grant number CE140100049.




\bibliographystyle{apacite}

\bibliography{RefPaper}

\begin{table}[htbp]
\caption{Results of inflection and outbreak size of each country/region}
\small
\begin{center}
\begin{tabular}{lccrrcccc}
\toprule
Country &  Model   & Infl.date & $n_{Infl}$ & $n_{\max}$   & $\phi_1$ & $\phi_2$ &  $\phi_3$  & $\phi_4$   \\ 
\midrule
Australia	&	1	&	29/03/2020	&	3727	&	7243	&	-394.78	&	7242.79	&	18.16	&	5.31	\\	
&	2	&	28/03/2020	&	3454	&	7245	&	-1.16	&	85.12	&	13.14	&	5.91	\\	
&	3	&	28/03/2020	&	3545	&	7174	&	1.68	&	3.86	&	8.73	&	5.12	\\	\hline
China	&	1	&	10/02/2020	&	45741	&	86985	&	-2076.92	&	86984.68	&	18.27	&	4.94	\\	
&	2	&	9/02/2020	&	42269	&	87225	&	-11.62	&	295.34	&	12.65	&	6.05	\\	
&	3	&	8/02/2020	&	38497	&	87498	&	0.11	&	4.94	&	-0.91	&	7.08	\\	\hline
France	&	1	&	5/04/2020	&	68129	&	151596	&	-15443.24	&	151595.90	&	26.99	&	10.32	\\	
&	2	&	4/04/2020	&	64046	&	151158	&	-41.37	&	388.79	&	17.64	&	10.80	\\	
&	3	&	4/04/2020	&	64615	&	151733	&	0.59	&	5.18	&	-2.02	&	11.52	\\	\hline
Germany	&	1	&	3/04/2020	&	80725	&	185753	&	-28519.04	&	185752.80	&	23.58	&	10.76	\\	
&	2	&	2/04/2020	&	75181	&	184438	&	-99.75	&	429.46	&	13.39	&	10.93	\\	
&	3	&	2/04/2020	&	77512	&	180651	&	-0.30	&	5.26	&	-3.91	&	10.16	\\	\hline
Italy	&	1	&	1/04/2020	&	106391	&	244335	&	-34334.78	&	244334.90	&	30.73	&	13.57	\\	
&	2	&	31/03/2020	&	101571	&	242952	&	-123.75	&	492.90	&	16.68	&	14.29	\\	
&	3	&	30/03/2020	&	98116	&	239062	&	-2.58	&	5.38	&	-12.65	&	14.00	\\	\hline
Russia	&	1	&	13/05/2020	&	240380	&	491381	&	-14640.86	&	491381.50	&	45.82	&	11.16	\\	
&	2	&	13/05/2020	&	240735	&	538392	&	-39.47	&	733.75	&	34.47	&	14.79	\\	
&	3	&	15/05/2020	&	260147	&	647295	&	-2.24	&	5.81	&	-13.28	&	20.69	\\	\hline
Spain	&	1	&	31/03/2020	&	104896	&	243948	&	-34451.39	&	243948.40	&	22.98	&	10.07	\\	
&	2	&	31/03/2020	&	104530	&	242397	&	-95.04	&	492.34	&	13.78	&	10.18	\\	
&	3	&	30/03/2020	&	98400	&	241883	&	-2.25	&	5.38	&	-8.91	&	10.59	\\	\hline
UK	&	1	&	24/04/2020	&	144057	&	306496	&	-28384.02	&	306496.40	&	39.13	&	14.63	\\	
&	2	&	21/04/2020	&	129028	&	309624	&	-132.79	&	556.44	&	21.37	&	17.10	\\	
&	3	&	18/04/2020	&	114591	&	298560	&	-5.79	&	5.48	&	-21.96	&	17.16	\\	\hline
USA	&	1	&	24/04/2020	&	905426	&	2123502	&	-375633.20	&	2123502.00	&	43.01	&	19.57	\\	
&	2	&	18/04/2020	&	727129	&	2357467	&	-3413.23	&	1535.40	&	-18.42	&	30.79	\\	
&	3	&	19/04/2020	&	767660	&	1957044	&	-7.53	&	6.29	&	-30.79	&	19.96	\\	\hline
Africa	&	1	&	7/06/2020	&	178552	&	357967	&	-2525.02	&	357966.50	&	76.83	&	18.50	\\	
&	2	&	13/09/2020	&	1708842	&	3953649	&	-128.87	&	1988.38	&	133.52	&	55.61	\\	
&	3	&	27/08/2020	&	967892	&	2606588	&	-93.15	&	6.42	&	-286.14	&	81.64	\\	\hline
Brazil	&	1	&	7/06/2020	&	665614	&	1315546	&	-2508.03	&	1315546.00	&	76.61	&	13.98	\\	
&	2	&	23/06/2020	&	1168848	&	2586527	&	-11.72	&	1608.27	&	75.27	&	24.31	\\	
&	3	&	21/07/2020	&	1960275	&	5240786	&	-46.63	&	6.72	&	-150.41	&	56.32	\\	\hline
India	&	1	&	17/06/2020	&	355594	&	709376	&	-6389.81	&	709375.50	&	78.61	&	17.14	\\	
&	2	&	26/07/2020	&	1088097	&	2510545	&	-83.42	&	1584.47	&	91.08	&	36.73	\\	
&	3	&	13/09/2020	&	2666949	&	7159223	&	-83.45	&	6.85	&	-260.10	&	79.90	\\	
\bottomrule
\end{tabular}
\label{tab:results}
\end{center}
\end{table}
   
\end{document}